\documentclass[10pt,twocolumn]{article}

\usepackage[utf8]{inputenc}
\usepackage[T1]{fontenc}

\usepackage{graphicx}
\usepackage{pdfpages}

\usepackage{ulem}

\usepackage{lipsum}
\usepackage{xcolor}

\usepackage{authblk}
\usepackage[margin=2cm]{geometry}
\usepackage{mathtools, cuted}

\usepackage{bm}
\usepackage{amsfonts}
\usepackage{amsmath}
\usepackage{amssymb}
\usepackage{mathrsfs} 
\usepackage{siunitx}

\title{Interface-templated crystal growth in sodium dodecyl sulfate solutions with NaCl}
\author[1,2]{Anna Kharlamova}
\author[1]{François Boulogne}
\author[2]{Philippe Fontaine}
\author[1]{Stéphan Rouzière}
\author[2]{Arnaud Hemmerle}
\author[2,3]{Michel Goldmann}
\author[1,*]{Anniina Salonen}
\affil[1]{Université Paris-Saclay, CNRS, Laboratoire de Physique des Solides, 91405, Orsay, France.}
\affil[2]{Synchrotron SOLEIL, L’Orme des Merisiers, Départementale 128, 91190, Saint-Aubin, France.}
\affil[3]{Institut des NanoSciences de Paris, Sorbonne Université, 4 Place Jussieu, 75005 Paris, France.}
\affil[*]{\small Corresponding author: anniina.salonen@universite-paris-saclay.fr}
\date{\today}
\begin{document}

\twocolumn[
    \begin{@twocolumnfalse}
        \maketitle
        \begin{abstract}
            Many ionic surfactants, such as  sodium dodecyl sulphate (SDS) crystallize out of solution if the temperature falls below the crystallization boundary. The crystallization temperature is impacted by solution properties, and can be decreased with the addition of salt. We have studied SDS crystallization at the liquid/vapor interfaces from solutions at high ionic strength (sodium chloride). We show that the surfactant crystals at the surface grow from adsorbed SDS molecules, as  evidenced by the preferential orientation of the crystals identified using grazing incidence X-ray diffraction. We find a unique timescale for the crystal growth from the evolution of structure, surface tension, and visual inspection, which can be controlled through varying the SDS or NaCl concentrations.
        \end{abstract}
        \vspace{2em}
    \end{@twocolumnfalse}
]

%
%

\section*{Introduction}

Many surfactants crystallize out of solution at low temperatures or in the presence of salts \cite{tadros2014introduction}.
This can be a problem for the stability of formulations, as crystallization can lead to a diminished visual aspect or a decreased foamability \cite{zhang2003defoammechanism}. However, solution crystallization is by no means always undesired.
It is an important step in a number of manufacturing processes \cite{HandbookIndustrialCrystallization,CrystallizaitonBook}, and surfactant crystals have been shown to act as effective foam stabilizers if they adsorb onto the bubble surfaces \cite{zhang2015precfoamAng, fameau2021aqueousreview}.

Solid particles have been used to create highly stable foams, through the practically irreversible adsorption of particles at the liquid/vapor interfaces (given suitable particle size and wetting conditions) \cite{rio2014unusually, binks2017colloidal}.
Such systems are particularly interesting for applications where foam lifetime is crucial.
We showed that the crystallization of sodium dodecyl sulphate (SDS) or potassium dodecyl sulphate (KDS) with the addition of either sodium or potassium chloride to SDS during foam generation leads to highly stable foams, which could be destroyed by heating them above the melting temperature of the crystals \cite{zhang2015precfoamAng}.
The criteria for foam stability have been explored, and the first requirements are a crystallization process which is fast compared to the foam destabilization processes, and a sufficient quantity of crystals \cite{zhang2017foams}.
Jiang \textit{et al.} \cite{jiang2020rolesalts} studied foam stability as different salts were added to SDS solutions (NaCl, KCl or CaCl$_2$).
They suggest a link between foam stability and crystal structure formed, with higher packing densities improving foam stability.
Binks and Shi \cite{binks2020aqueousMg} used magnesium nitrate hexahydrate to crystallize the solutions of SDS.
They saw a correlation between the size of crystals, foamability and foam stability, with the intermediate size of crystals being the most efficient.
This was in agreement with suggestions when a series of alkali salts had been used for the precipitation \cite{zhang2019JSurfDetergAlkalis}.

Although the crystal structure, size of crystals and the rate of crystallization have been identified as important parameters in the capacity of the crystals to stabilize foams, there are few studies of SDS crystallization at liquid/vapor surfaces.
In the bulk the crystallization of SDS has been studied, and it is known that a number of crystal structures can be formed, depending on the degree of hydration \cite{SMITH2000173, sundell_crystal_1977,KEKICHEFF1989112}. Howevever, as recently as in 2021, a novel hydrate form of SDS crystal was discovered \cite{lee2021novel}. Recently, Khodaparast \textit{et al.} explored the impacts of confinement and surfaces on the crystallization of SDS \cite{khodaparast2020surface}.
The use of microvolumes allowed them to study the impact of surface roughness and surface energy on crystallization. They showed that increasing the surface energy decreased nucleation times.

The crystallization of surfactants is different from the \textit{surface freezing} phenomenon. Long-chain alkanes have been shown to undergo a surface freezing transition, where at temperatures above bulk freezing they form a nanometric crystal-layer at alkane/water interfaces \cite{Wu_Ocko_Alkanes_1993,Ocko_Wu_PRE_1997}. Surface freezing has also been used to shape emulsion drops \cite{DenkovNatureSurfaceFreeze,Guttman_emulsion_drops}. Such a transition has been shown to occur at gas/oil interface in the presence of a long-chain surfactant \cite{Lei_Bain_surfinduced}. In surface freezing the systems are kept above bulk crystallization temperature, however the role of surfaces in the crystallization of surfactants is still an open question.

In order to study SDS crystallization at liquid/vapor surfaces, we have used varying concentrations of NaCl to modulate the temperature of crystallization and hence the kinetics of crystal formation.
The addition of salt screens the electrostatic repulsion between the surfactant headgroups and promotes precipitation. This increases the crystallization temperature, so that crystals form at room temperature.
We measure the crystal structure at the surface using grazing incidence X-ray scattering, and follow the kinetics of crystallization with surface tension and imaging methods.

%
%

\section*{Materials and methods}
\subsection*{Materials}
The surfactant sodium dodecyl sulfate (SDS, purity $\ge 99.0\%$) and the salt sodium chloride (NaCl, purity $\ge 99.0\%$) are purchased from Sigma.
The SDS powder is used as received and the NaCl powder is roasted in an oven at 650--700~$^{\circ}$C for 16-20 hours to remove any possible organic impurities.

\subsection*{Preparation of SDS solutions with NaCl}
Solutions are prepared in the concentration range 0.5--6.0 mM for SDS and 350--600 mM for NaCl volumetrically at 20~$^{\circ}$C. These concentrations were chosen to observe crystallization on reasonable timescales.
Required amounts of SDS and NaCl powders are transferred into a 500 mL or 1 L flask and diluted with ultrapure water (18.2 M$\Omega\cdot$cm).
The precipitate forming in the solutions during preparation is solubilized by increasing the temperature to 25--30~$^{\circ}$C (depending on the NaCl concentration) in a water bath (Memmert WNE 10 equipped with a cooling Peltier).
The temperature is then gradually decreased back to 20~$^{\circ}$C, which requires 30--40 minutes.
All prepared solutions were homogeneous and transparent prior to the start of the experiments, and are analyzed immediately after preparation.

\subsection*{The Krafft boundary determination}
The Krafft boundary is determined for solutions with [SDS] = 1, 5, 10 and 100 mM and [NaCl] from 50 to 1000 mM.
The mixtures with different surfactant/salt concentrations are prepared from stock solutions of SDS and NaCl in 20 mL glass bottles.
The mixtures are heated at approximately 40~$^{\circ}$C until the precipitate is solubilized to ensure homogeneity and placed in a fridge at 4~$^{\circ}$C overnight.
The bottles are then gradually heated in a water bath (VWR 1137-1P circulating water bath) starting from 4 $^{\circ}$C with the rate 1 $^{\circ}$C per 2 hours.
The Krafft boundary is considered as the lowest temperature at which all precipitate  appears dissolved by visual inspection.
We estimate the uncertainty of the melting temperature measurement in $\pm$1 $^{\circ}$C.
The measured temperatures are shown in Supporting Information Figure S1, and all the data points are shown in Table S1. The measured values are in good agreement with the study of Illous \textit{et al.} \cite{Illous_SDS_NaCl_Krafft}.
The Krafft boundary does not depend on the concentration of SDS, but the temperature increases weakly with [NaCl]. With 100 mM NaCl the crystals melt at 20~$^{\circ}$C, while at 1~000~mM the melting temperature is around 32~$^{\circ}$C.

\subsection*{Crystallization kinetics}
The kinetics of crystallization at the interface of salt-surfactant mixtures is characterized using a polytetrafluoroethylene Langmuir trough (Kibron microtrough) 80 mm $\times$ 206 mm equipped with two polyoxymethylene barriers (width 20 mm) that are immobile during the measurement.
The surface area of the trough is therefore 132.8 cm$^2$.
The trough is equipped with a balance holding a sensor -- the DyneProbe, made of a metal alloy with a hydrophilic oxide, ensuring negligible contact angle with the solution.
The trough operates in an acrylic cover  box preventing perturbations during measurements.
The temperature in the trough  is regulated with a water bath at 20 $^{\circ}$C.
The sensor is cleaned from organic residues by heating with a flame torch and calibrated with 100 mL of ultrapure water at 20 °C before each measurement.
Then 100 mL of freshly prepared SDS + NaCl solution at 20 $^{\circ}$C is poured into the trough and the time evolution of the surface tension is recorded.
The surface tension is initially stable and is followed by a sharp drop after crystallization at the interface. For each measurement, we determined the characteristic time of the surface tension drop $\tau_{\gamma}$ as the inflection point between the plateau and the slope.

To correlate the percolation of crystals at the interface with the surface tension measurements, crystallization kinetics for solutions with 500 mM NaCl was also followed by taking photographs of the solution interface in a glass Petri dish (diameter: 11 cm, surface area 95 cm$^2$) placed in the Kibron trough. The Petri dish was colored with black paint on the outside for easier observation of the crystallization process. The trough was filled with water at 20 °C that surrounded the bottom of the dish for temperature regulation. The DyneProbe, cleaned with a flame torch, was calibrated with 100 mL of ultrapure water poured into the dish that was then substituted with 100 mL of the SDS + NaCl solutions prepared as described above. The surface tension was measured as function of time and grayscale photographs of the solution interface were taken every minute with a Basler acA3800-14uc USB 3.0 camera placed at approximately 45° angle with the liquid interface.
The percolation time $\tau_{perc}$ is determined from the movies by eye as the time at which the movement at the interface stopped due to its complete coating with crystals. We estimate the uncertainty of the values to be $\pm10$ min. We verified with image analysis in MATLAB that the values determined by the eye correspond to the time to reach the plateau of total pixel intensity at the water interface (the result is not shown).

To obtain colorful images of the evolution of the crystal growth, we filmed solutions (100 mL) in an open Petri dish (diameter: 11 cm), colored black from the outside, in a lab with air temperature control (20 °C). The images were taken every minute with a USB 2.0 camera (uEye UI-1495LE-C-HQ).

\subsection*{X-ray Scattering}

X-ray scattering experiments are carried out at the SIRIUS beamline at SOLEIL Synchrotron (Saint-Aubin, France).
Details of the beamline optics can be found in the literature \cite{Hemmerle2024}.
The X-ray beam energy is fixed by a Silicon (111) Double Crystal Monochromator (DCM) at 8 keV ($\lambda = 0.155$~nm).
The beam size is fixed horizontally by optical slits and vertically by a focalizing mirror at $0.1 \times 0.5$~mm$^2$ (V $\times$ H) for Grazing Incidence X-ray Diffraction (GIXD) experiments and $0.1 \times 2$~mm$^2$ for X-ray diffraction measurements in the vertical plane of incidence.

\paragraph{GIXD experiments}

For Grazing Incidence X-ray Diffraction (GIXD) experiments, the water
surface is illuminated at an incident angle of $2$~mrad below the
critical angle ($2.7$~mrad) of total external reflection on the air/water interface so that the incident wave is almost totally reflected.
In this regime, the refracted wave is evanescent, probing the interface with a penetration depth of a few nanometers.
The diffracted X-ray beam intensity is recorded by a Pilatus3 1M 2D detector (Dectris, Switzerland) as a function of the horizontal in-plane scattering angle $2\theta$ between the incident beam and the scattered beam.
The in-plane resolution ($\Delta 2 \theta = 0.06^o$) is defined by a Soller-Collimator (JJ X-ray, Denmark) located in front of the Pilatus3 1M detector.
A home-made Langmuir trough of total area 800 cm$^2$ and with 400 mL volume is used for these surface crystallization experiment combined with X-ray measurements.
It is enclosed in a temperature-controlled sealed chamber and flushed with helium gas during data collection in order to lower scattering and sample damage.
The helium is saturated with water before entering the trough enclosure in order to avoid water evaporation during the measurements.
The surface pressure is measured in this trough by the Wilhelmy plate method, using a filter paper plate and a Riegler and Kirstein Gmbh (Berlin, Germany) surface pressure sensor.

\paragraph{Diffraction in the vertical plane of incidence}
Out-of-plane diffraction (\textit{i.e.} in the plane containing the incident beam and its specular reflection) measurements were performed in an original geometry initially developed for X-ray reflectivity at liquid-air interfaces. In this setup, the incidence angle on the last deflecting mirror of the SIRIUS beamline is varied to scan the incidence angle on the water surface between 0 and 2$^\circ$.
The upper limit of 2$^\circ$ on the water surface is determined by the critical angle of total external reflection of the mirror of approximately 0.5$^\circ$, given by its external coating of 80 nm platinum.
When the incidence angle on the mirror exceeds this threshold, the intensity of the reflected X-rays decreases significantly, with almost no X-rays being reflected for angles greater than 1$^\circ$ on the mirror, equivalent to 2$^\circ$ on the water surface. The intensity of the incident beam $I_0$ is monitored by an ionization chamber (IC PLUS 50, FMB Oxford, England) placed right before the liquid surface for proper normalization of the reflected beam.
In this particular geometry, as the incident angle increases, it becomes necessary to lower the liquid surface to follow the incident beam. This adjustment is achieved by vertically translating the sample stage of the diffractometer, which has a wide range of motion ($150$~mm) and an excellent repeatability ($1$~$\mu $m). The signal reflected by the interface is measured by the Pilatus3 1M detector, which is moved vertically to ensure that the reflected spot remains at the same position on the detector.
For every data point of the experimental curve, the signal is measured by summing the intensity of each pixel within a Region Of Interest (ROI) that is centered around the reflected beam. Background subtraction is then performed by calculating the average signal from the ROIs located immediately on the left and on the right of the ROI containing the reflected beam.
Finally, the signal is divided by a measurement of the direct beam on the detector, along with a proper normalization by the $I_0$ intensity.

%
%

\section*{Results and discussion}

\begin{figure*} 
    \includegraphics[width=\textwidth]{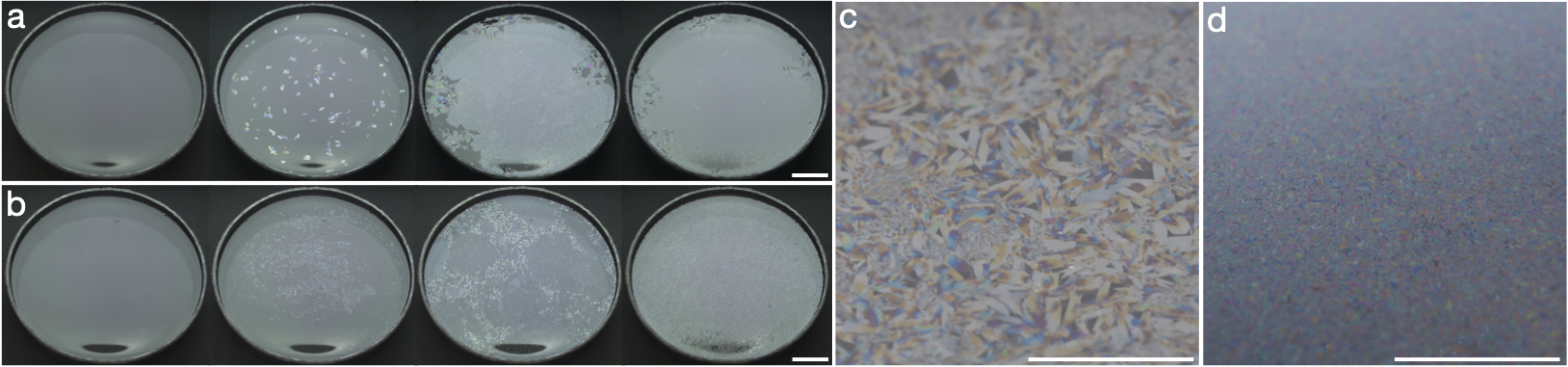}
    \centering
    \caption{Time series of the surface of a glass Petri dish with an SDS-NaCl mixture at 20~$^\circ$C with (a) 500 mM NaCl and 0.6 mM SDS at 0, 8, 13.5 and 20 hours and (b) 500 mM NaCl and 1 mM SDS at 0, 1, 2 and 5 hours. (c) Close up of the crystal surface with 500 mM NaCl and 0.6 mM SDS after 15 hours. (d) Close up of the crystal surface with 500 mM NaCl and 1 mM SDS after 15 hours. The scale bar is 2 cm in all the photographs.
    See movies in Supplementary materials.
    }
    \label{fig:visualization}
\end{figure*}

\subsection*{Macroscopic observation of the surfaces}

We observe the evolution of solutions of SDS in Petri dishes. At the SDS concentrations studied, and at room temperature, the surfactant will not precipitate without the addition of salt. The addition of 500 mM NaCl will decrease the solubility of SDS and the critical micelle concentration (CMC) falls to around 0.5 mM \cite{Phillips_SDS_NaCl_CMC}.
We are working at the concentrations above this concentration. We observe crystal-formation, as the melting temperature of SDS crystals is around 25 $^\circ$C as shown the Supporting Information Figure S1.
We have selected images of the time-evolution of the surface of a Petri dish with two different solutions at 0.6 mM (Figure \ref{fig:visualization}a) and 1.0 mM (Figure \ref{fig:visualization}b) SDS solution with 500 mM of NaCl.
The videos from which the images are extracted are available as supplementary materials.
In both samples the surface is free from any visible crystals after pouring the solution in the Petri dish. Eventually, crystals will appear in both samples, however, the timescales of the process are different.

For the sample with 0.6 mM SDS (Figure \ref{fig:visualization}a), we need to wait several hours before the appearance of crystals (the second photograph of the series was taken 8 hours after preparation).
The crystals formed are large, several millimeters in size and with varying colours. This means that they have a typical thickness of hundreds of nanometers at this moment.
Over several hours, the crystals continue to grow in number and in size until reaching an almost full coverage of the Petri dish.
Their motion ceases as they percolate through the surface.  At longer times the visual aspect changes very little, although it continues to become more homogeneous in hue.

Increasing the concentration of SDS to 1.0 mM (Figure \ref{fig:visualization}b) leads to a much faster formation of crystals, which are also much smaller (millimetric).
After 1 hour (2nd photograph) large numbers of sub-millimetric crystals can be seen at the sample surface.
These continue to grow in number, and in size until they cover the full surface.
As the crystal layer continues to thicken, the brightly colored crystals thicken and the film becomes colorless.

In both cases, the surfaces are very mobile during the growth of the crystals, due to convection, however, once the surface density of crystals becomes sufficiently high all motion arrests and the surfaces become jammed. This is not the end of the evolution, and the thickness of the layers continues to increase.

The impact of the SDS concentration can also be seen in photographs of a smaller region of the surfaces, as shown in Figure \ref{fig:visualization}c and d, for the 0.6 mM and 1.0 mM SDS respectively, both taken after 15 hours.
We can again note the difference in the size of the crystals, as those formed in the 0.6 mM SDS are much larger than with 1.0 mM SDS. This is typical of the samples, where smaller SDS concentrations form larger crystals. We have not explored the control over the crystal size, however the use of salts has been shown to be an interesting way to control crystal size \cite{Milleretal_CrystalSize}.

\subsection*{Crystal structure at the surfaces}

\begin{figure}
    \centering
    \includegraphics[width=\linewidth]{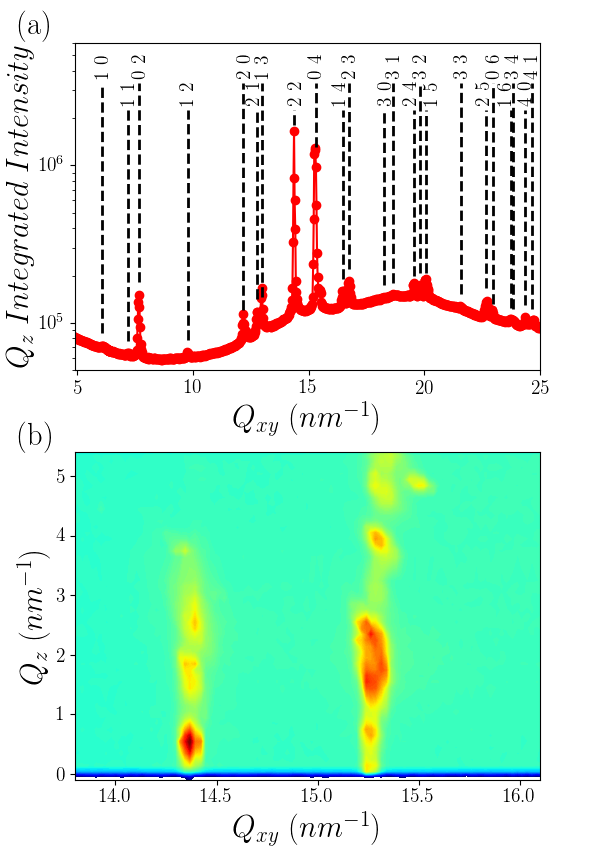}
    \caption{(a) $Q_z$-Integrated Grazing Incidence X-ray Diffraction spectra measured on a $1.5$ mM SDS and $500$ mM NaCl solution  12 hours after the SDS:NaCl solution was poured into the trough. This time frame corresponds to a later stage, following the complete coverage of the surface by crystals. Peaks are indexed by a rectangular 2D structure with $a=1.032$~nm and $b=1.641$~nm.
    (b)  $Q_{xy}$--$Q_z$ intensity map of diffraction peaks $22$ and $04$.}
    \label{fig:structure1}
\end{figure}

In order to determine the structure and, eventually, the nature of the crystals that form at the surface of SDS solution in presence of high concentration of NaCl, we performed GIXD and diffraction in the vertical plane of incidence.
Figure~\ref{fig:structure1}-a presents the $Q_z$--integrated diffraction  spectrum measured after the complete covering of the free surface of an SDS solution (1.5 mM) and NaCl (500 mM).
A very large $Q_{xy}$--range is scanned and revealed a high number of diffraction peaks.
The $Q_{xy}-Q_z$ intensity map (Figure~\ref{fig:structure1}-b) clearly demonstrates  that the vertical structure of the diffraction peaks corresponds to modulated diffraction rods indicating that the crystals are oriented with a defined lattice plane always parallel to the surface.
Since the crystals are oriented by the interface, it also suggests that the crystal growth occurs at the solution-air interface.
In Figure~\ref{fig:structure1}-a all diffraction peaks are indexed on a two-dimensional rectangular lattice with $a=1.032$~nm and $b=1.641$~nm.
To obtain this lattice, we started from the structures of SDS powder proposed by Smith \textit{et al.} \cite{SMITH2000173}, who determine the crystallographic structure of SDS crystals with different degrees of hydration.
The structure is always composed of lamellae of SDS molecules crystallizing in bilayers formed by alternate layers.
We calculate the peak position for each structure proposed by Smith \textit{et al.}, with different amounts of water in the unit cell, using the xrayutilities python library \cite{Kriegner:rg5038}.
To compare with our 2D diffraction peaks, $\mathbf{a}$ and $\mathbf{b}$ vectors were ascribed to the 2D structure parallel to the lamellae, and the $\mathbf{c}$ -- to the pile-up of the lamellae. We then only keep the $hk0$ peaks to compare with the experimental peak positions.
Among the different hydration levels of the SDS crystals, the ${\rm SDS}: 1/8~{\rm H}_2{\rm O}$ gives parameters $a=1.022$~nm and $b=1.641$~nm that allow a very good adjustment to the peaks in Figure~\ref{fig:structure1}-a.
Only a small variation of $a$ to 1.032 nm is necessary to fully reproduce the measured peak positions.
Several reasons may explain this difference: the experimental resolution of the instrument is about 0.05 nm$^{-1}$ and leads to an uncertainty in distance determination roughly equivalent to the measured difference; also the hydration level is neither controlled and nor measured and could be slightly different from the crystals of Smith \textit{et al.} \cite{SMITH2000173}.

\begin{figure}
    \centering
    \includegraphics[width=\linewidth]{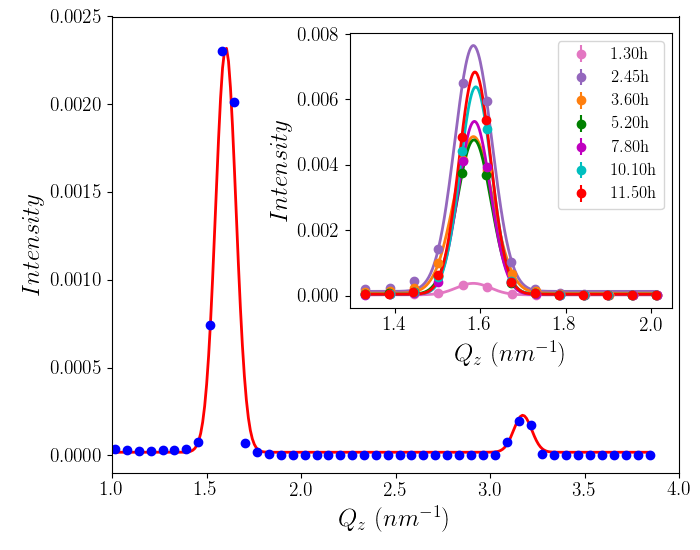}
    \caption{X-ray diffraction measured in the plane of incidence with the X-ray Reflectivity geometry on a solution of SDS 1.5~mM SDS and 500 mM NaCl.
    Inset: Intensity as function of time expressed in hours in the legend in the $Q_z$-range of the (001) peak for a solution of 1.0~mM SDS and 500~mM NaCl. Lines are a Gaussian fit of the experimental data.
    }
    \label{fig:structure2}
\end{figure}

In order to determine the vertical structure and the distance between lamellae we performed  X-ray diffraction in the vertical plane of incidence using the same experimental geometry as X-ray reflectivity measurement (XRR).
The main feature of the diffraction spectra in the vertical plane is the presence of two Bragg peaks as shown in Figure~\ref{fig:structure2} located at $q_1=1.603 \pm 0.002$ nm$^{-1}$ and $q_2=3.172 \pm 0.022$ nm$^{-1}$.
The ratio of these two positions is about two and can thus be attributed to the diffraction peaks of a multi-layer with a period of $2 \pi / q_1 = 3.9$~nm.
In the inset of figure~\ref{fig:structure2}, we have plotted the first order peak (001) measured with time.
Its intensity is growing while the other parameters remain constant, showing that the amount of crystallized matter is increasing with time.
The Gaussian shape of the diffraction peaks suggest that they are limited by the instrument resolution. The $w=0.07~{\rm nm}^{-1}$ width of the diffraction peaks is indeed around the expected experimental resolution. This suggests that the correlation length $\xi$ is more than $80 {\rm nm}$ in the vertical direction according to the Debye-Scherrer formulae ($\xi \approx 0.9 \times 2 \pi / w)$ \cite{Barton1998, kaganer1999}.

The X-ray diffraction studies of the crystalline layer formed at the free surface of SDS/NaCl solutions with time demonstrates that the formed structure corresponds to hydrated SDS crystals oriented by the interface since the SDS lamellae formed are parallel to the surface.
Comparing the structure obtained with literature data\cite{SMITH2000173}, the in-plane rectangular structure and its parameters $a$ and $b$ compared well with the lattice parameters $b=1.02$~nm and $c=1.64$~nm and the $\alpha=90^\circ$  angle of the 2D structures of the lamellae of the ${\rm SDS}: 1/8~{\rm H}_2{\rm O}$ hydrated structure evidenced by Sundell et al. \cite{sundell_crystal_1977}. For the third parameter corresponding to the lamellar periodicity, the $3.9$~nm periodicity we found is exactly half of the $a=7.8$~nm lattice parameter of the ${\rm SDS}: 1/8~{\rm H}_2{\rm O}$ hydrated structure \cite{sundell_crystal_1977, SMITH2000173}. In both papers, Sundell et al. and Smith et al. pointed out the fact that for this hydration ratio, the unit cell contains 32 SDS molecules and 4 molecules of water, resulting in two bilayers in the unit along the axis perpendicular to the lamellae~\cite{sundell_crystal_1977, SMITH2000173}.
This double bilayer cell is mandatory to account for an alternating  displacement of adjacent molecule perpendicular to the layer place \cite{sundell_crystal_1977}. However we do not have enough measured peaks in the limited Qz-range explored to perform such a refined analysis of the structure. Therefore we have simply indexed the peaks $q_1$  and $q_2$ as (001) and (002) with $c=3.9$~nm, but they could also be indexed (002) and (004) with the double lattice parameter $c=7.8$~nm.
We can conclude that the formed crystals at the concentrated salt / SDS solution interface are hydrated SDS crystals with a proportion of $1/8$~${\rm H}_2{\rm O}$ molecules.

 Apart from anhydrous SDS crystals, several hydrates have been identified in the water / SDS phase diagram in the literature: ${\rm SDS}: 1/8~{\rm H}_2{\rm O}$ \cite{sundell_crystal_1977}, ${\rm SDS} : 1/2\ {\rm H}_2{\rm O}$, monohydrate \cite{SMITH2000173}, dihydrate \cite{KEKICHEFF1989112}.
 The water molecules exhibit strong interactions with the polar head groups of SDS molecules and are located between the lamellae in the lamellar phase \cite{SMITH2000173, KEKICHEFF1989112}.
 In the SDS / water  phase diagram, the ${\rm SDS}: 1/8~{\rm H}_2{\rm O}$ hydrate appears at low water fractions (above 90wt \% SDS concentration) \cite{KEKICHEFF1989112}. At first glance, it may seem surprising to discover the hydrate with the least amount of water in a system that contains a large reservoir of water.
 However, we should remember that we are working at high salt concentration (500 mM NaCl). The SDS hydrate structure can be modified through changing solvent conditions, such as using acetone as an antisolvent to produce ${\rm SDS}: 1/8~{\rm H}_2{\rm O}$ crystals \cite{Lee_antisolvent_structures}. Indeed, we show in Fig.~S2 of the Supplementary materials that the lamellar spacing of the bulk crystals shifts in the presence of NaCl.
 The lamellar spacing measured with 600 mM NaCl is 3.9 nm in agreement with the structure measured at the surface.

\subsection*{Crystal growth kinetics}

\begin{figure}[ht]
    \centering
    \includegraphics[width=\linewidth]{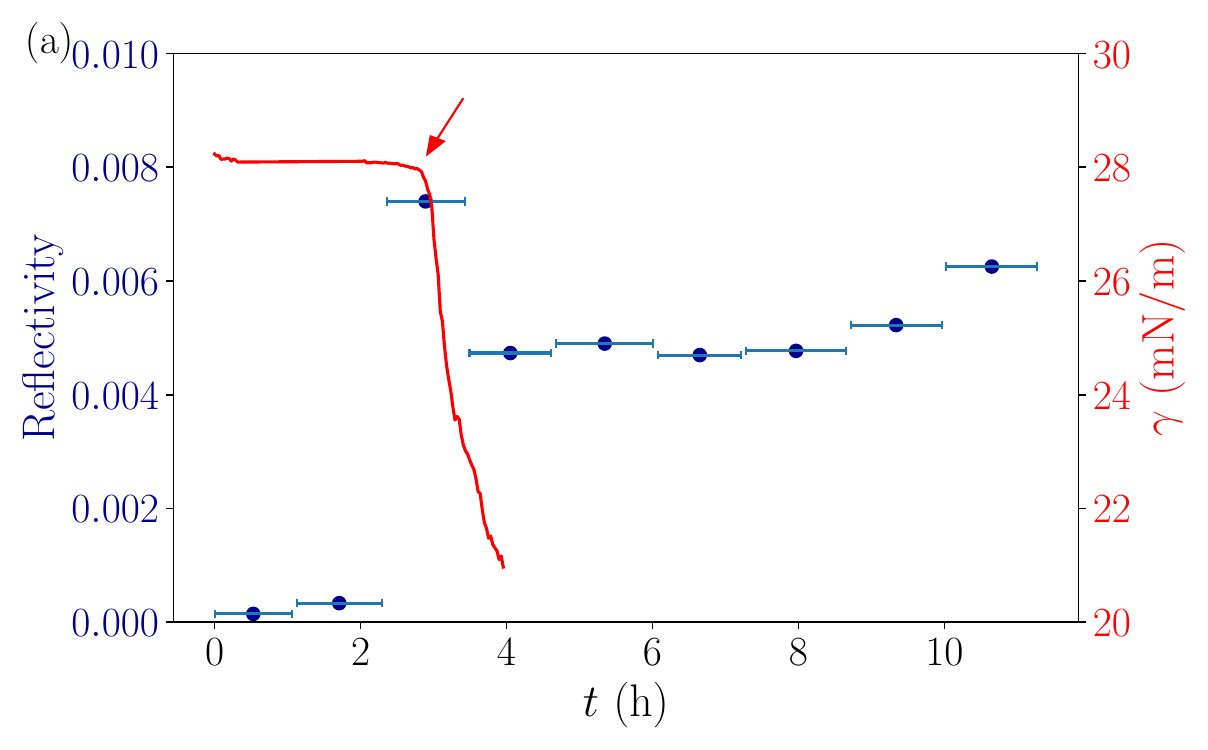}\\
    \includegraphics[width=\linewidth]{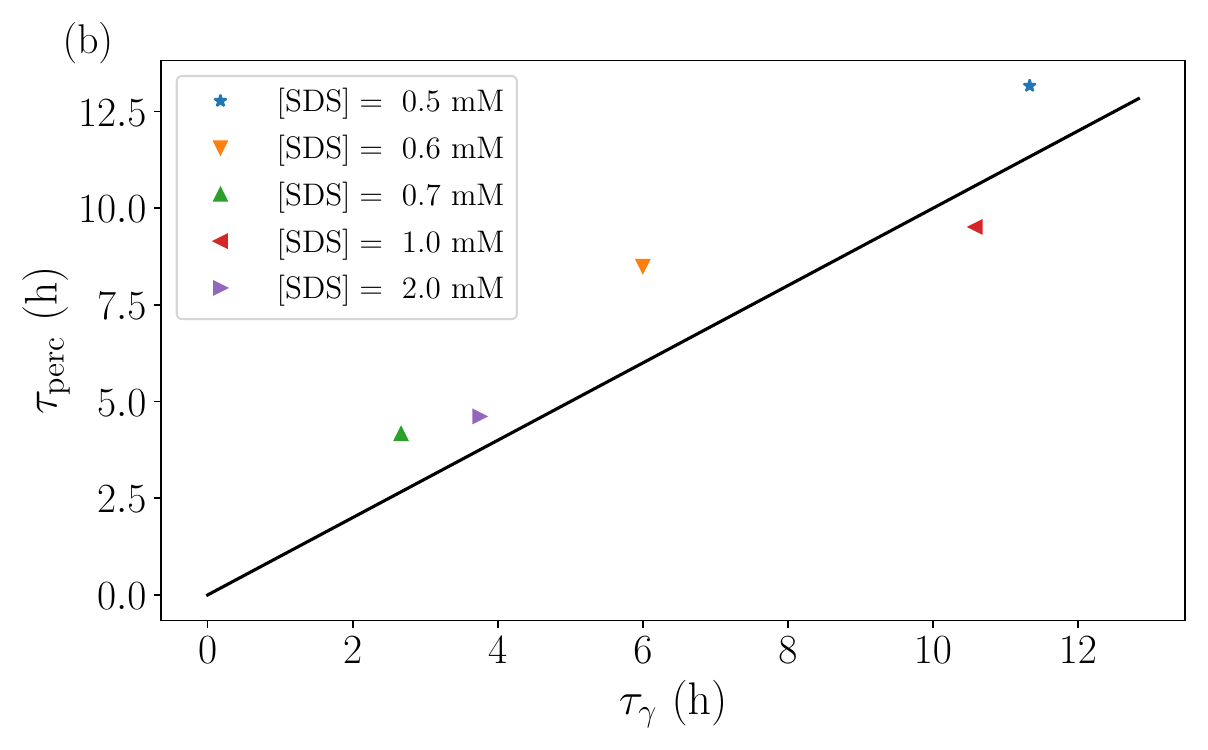}
    \caption{(a) Amplitude of the (001) diffraction peaks obtained by fitting the intensity of Figure~\ref{fig:structure2}-Inset by a Gaussian function as function of time for a $1.0$~mM SDS and $500$~mM NaCl solution.
    The horizontal bars indicate the time over which the measurement is performed.
    The red line shows the simultaneous measurement of the surface tension $\gamma$ for which the time $\tau_{\gamma}$ is indicated by a read arrow.
    (b) Percolation time $\tau_{\rm perc}$  as a function of $\tau_\gamma$ for five different SDS concentrations and $[{\rm NaCl}]=500$~mM.
    The black solid line represents equality between axes.
    }
    \label{fig:durations}
\end{figure}

 To have a more quantitative description of the interface-templated crystal growth phenomenon, we have combined the visual inspection of the surface, the measurement of the time-evolution of the surface tension $\gamma$, and the intensity of the (001) peak obtained in diffraction in the vertical plane of incidence.
 Results obtained for the same SDS and NaCl solution as in Figure~\ref{fig:visualization} are plotted in Figure~\ref{fig:durations}-a.
The time to measure a single spectrum of diffraction in the vertical plane of incidence is about 45 min, which explains the low time resolution in Figure~\ref{fig:durations}-a. However, the depicted evolution is typical of the studied samples.
It initiates with the appearance of a peak with low intensity. Around the percolation time $\tau_{\rm perc}$, the intensity of the (001) peak significantly increases and subsequently decreases to approximately 60~\% of its maximum value.
It remains stable for a period and then shows a further increase over longer times.
While the increase at longer times can be attributed to the thickening of the interfacial layer, explaining the peak intensity maximum at the percolation of the layer poses a greater challenge.

One possible explanation could be sample degradation, as the crystals at the interface undergo percolation and the two-dimensional diffusion is halted, resulting in continuous irradiation of the same crystals. This phenomenon may induce irradiation damage, causing a decrease in peak intensities. However, typically, this would also result in an increase in peak width, which is not observed in this case. Furthermore, the subsequent increase in intensity over longer times does not support this explanation.

An alternative explanation could arise from the observation that at the percolation time, the surface is relatively flat, and the crystals align well with the solution's surface. This ideal scenario facilitates the measurement of diffraction from a two-dimensional, surface-oriented powder. However, after percolation, the roughness of the surface increases, and crystals may become misaligned, resulting in a smaller diffraction signal.

The evolution of the reflected intensity is mirrored in the response of the surface tension measurements. Initially, the surface tension is about $31.5~\textrm{mN}\cdot\textrm{m}^{-1}$ far below the surface tension of ultrapure water ($72~\textrm{mN}\cdot\textrm{m}^{-1}$ at 20~$^\circ$C).
At a given time $\tau_{\gamma} \approx 3$~h, a decrease followed by a sharper drop of the surface tension occurs concomitantly with the end of the crystal expansion at the surface as illustrated in Figure~\ref{fig:visualization}.
Indeed, the measurement of the surface tension, performed by a Wilhelmy plate, is unreliable when the liquid surface becomes rough, as expected when completely covered by solids domains.
Thus, surface tension values obtained after $t=\tau_{\gamma}$ are not usable.
Data collecting continued after this time to allow us to accurately determine $\tau_\gamma$, Figure~\ref{fig:durations}a.

In addition, the diffraction measurements indicate an absence of the (001) peak just after the sample preparation, supporting the absence of crystals at the interface.
Then, the peak intensity  increases with the growth of crystals and reaches a maximum correlated with the decrease of the surface tension.
After this time, no drastic change of the intensity is measured.
Even if visual observations evidence a thickening of the crystalline layer, GIXD cannot measure it due to the limited penetration of X-rays for thick layers, only the diffraction in the vertical plane of incidence is still sensitive at long time.

The temporal evolution of the intensity is typical for all the samples with different SDS concentrations. They show the appearance of an intense diffraction peak at percolation. The peak intensity decreases just after, followed by a slow increase at longer times.

The characteristic time-scales linked to structural evolution and surface tension measurements are in close agreement (Figure \ref{fig:durations}a).
From the images taken of the interfaces, as shown in Figure \ref{fig:visualization} we can also fix a characteristic time $\tau_{\rm perc}$ defined as the time for the crystals to percolate through the interface.
At $\tau_{\rm perc}$ the highly mobile surface suddenly slows down as the crystal coverage become sufficiently high to block the mobility.
In Figure \ref{fig:durations}b a comparison between $\tau_{\rm perc}$ and $\tau_{\gamma}$ is shown. The equivalence between the times indicates that we can define a single characteristic time-scale for the development of the surface crystals at the vapor/water interface, which gives rise to a response in surface tension, local structure and macroscopic appearance.

\begin{figure}
    \centering
       \includegraphics[width=\linewidth]{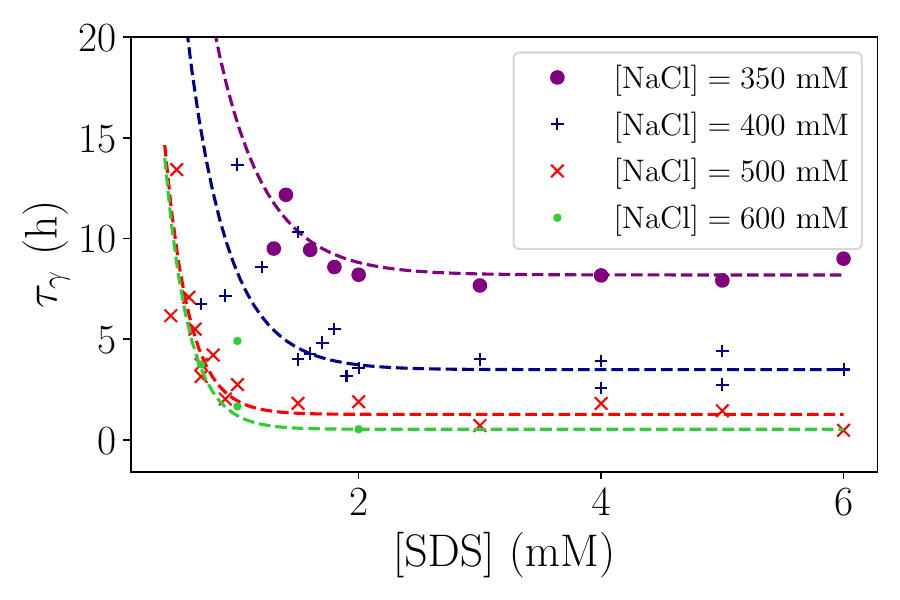}
       \caption{Surface tension drop time $\tau_{\gamma}$ as a function of SDS concentration for four NaCl concentrations.
       Lines are guides for the eye.}
    \label{fig:durations-with-conc}
\end{figure}

Then, we focus on the effect of salt and surfactant concentrations on the crystal growth kinetics.
In Figure~\ref{fig:durations-with-conc}, we have plotted the time $\tau_{\gamma}$ based on the surface tension measurements  as a function of the SDS concentration for different salt concentrations.
For each salt concentration, the evolution is characterized by a sharp decrease as SDS concentration increases followed by a constant time $\tau_{\gamma}$ at high surfactant concentration that increases with the reduction of the amount of SDS in the solution.
As the salt concentration increases from 350 to 600 mM the melting temperature of the crystals increases from 23 to 26$^\circ$C, and the CMC decreases from 0.6 to 0.4 mM \cite{Phillips_SDS_NaCl_CMC,Naskar_CMC_SDS_NaCl}. This leads to a higher degree of supersaturation, which will increase the rate of crystal formation. However, an increase in the NaCl concentration also leads to a higher surfactant surface coverage as the repulsion between the surfactant headgroups is screened. Both of these effects can increase the kinetics of crystallization at the surface.

%
%
\section*{Conclusion}

We have studied the crystallization of SDS at the vapor/water interface as we precipitate the surfactant from salt solution (NaCl). We measure the structure of the surface layers using a combination of GIXD and diffraction in the vertical plane of incidence, to access both the in-plane and out-of-plane structure.
We show that crystals are formed from adsorbed SDS molecules, and that they are oriented parallel to the surface.
We can match the crystal structure to that of a poorly hydrated SDS (${\rm SDS}: 1/8~{\rm H}_2{\rm O}$ hydrate), despite the dilute conditions and contact with a water bath.
This is because the presence of salt changes the equilibrium bulk crystal structure to a less hydrated form.

We analyze the kinetics of crystal growth by combining the structural evolution with surface tension measurements and visual observations. We show that a single characteristic time  defines the evolution of the crystals.
The visual percolation of the surface is accompanied by a sudden decrease of the measured surface tension, and an increase of the (001) peak intensity.

The characteristic time for the precipitation of SDS is long compared to the characteristic timescales linked to foam stability (hours compared to tens of minutes). This is probably why the use of NaCl to precipitate SDS for the generation of stable foams is less frequent, and potassium or magnesium chlorides are much more efficient\cite{zhang2015precfoamAng, binks2020aqueousMg}. However, this system shows variation in crystal size, which would be interesting to explore and to link to bubble stability.

\section*{Supporting Information}
Two videos of the surface crystallization. The data on the Krafft boundary. Bulk wide-angle X-ray scattering spectra of the precipitate.

\section*{Acknowledgments}
We thank Fabrice Cousin, François Muller, Patrick Kékicheff and Matthieu Roché for fruitful discussions.
We also thank Zhizhong Li for his assistance on salt purification.
We gratefully acknowledge SOLEIL synchrotron for provision of beamtime on SIRIUS Beamline. This work is supported by “Investissements d’Avenir” LabEx PALM (ANR-10-LABX-0039-PALM) project "Interfreeze".

\bibliography{biblio}

\bibliographystyle{vancouver}

\newpage\clearpage
\includepdf[pages={1-}]{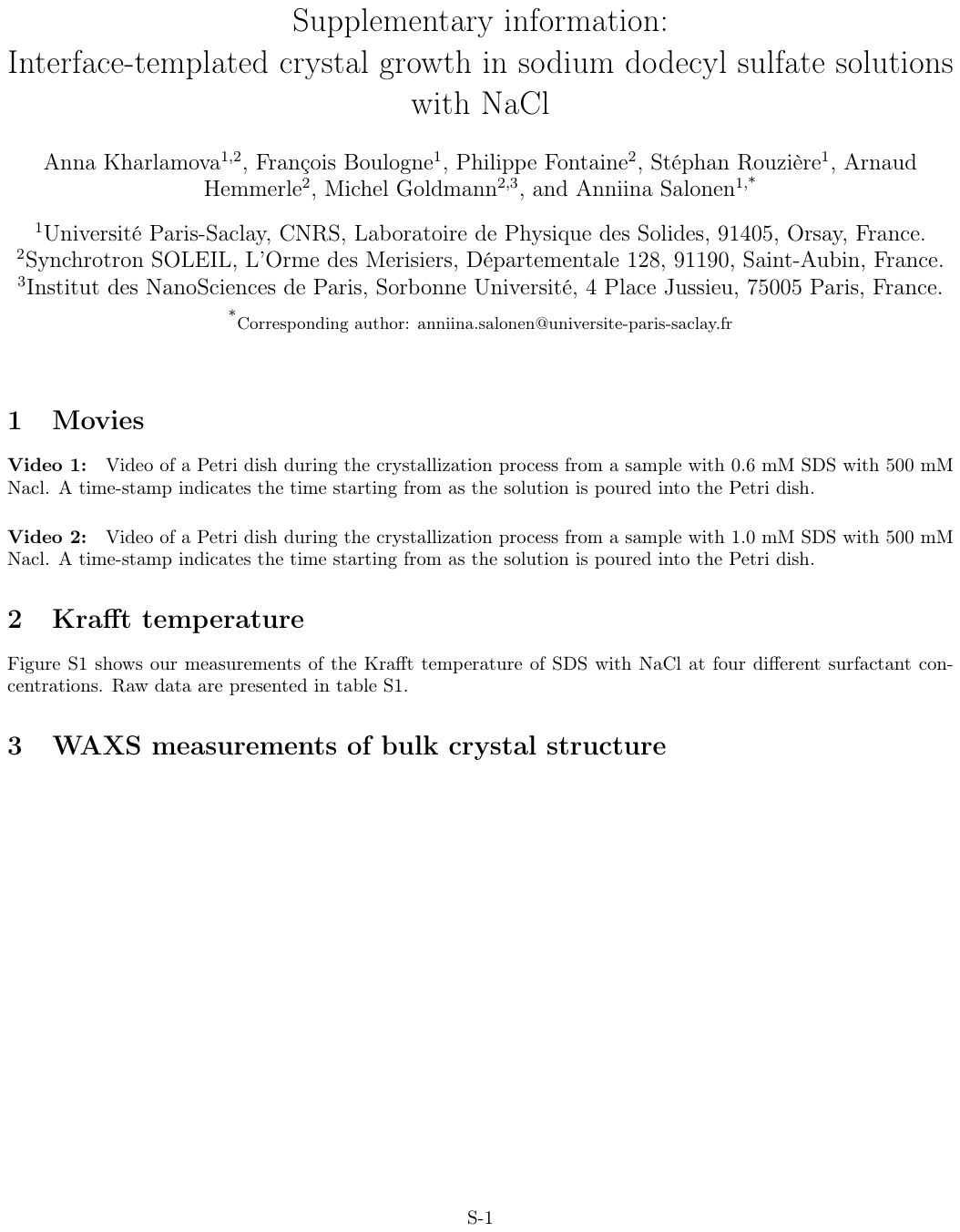}

\end{document}